# Collapse of Primordial Gas Clouds and the Formation of Quasar Black Holes

Abraham Loeb[1] and Frederic A. Rasio[2]

Institute for Advanced Study, Olden Lane, Princeton, NJ 08540

## ABSTRACT

The formation of quasar black holes during the hydrodynamic collapse of protogalactic gas clouds is discussed. The dissipational collapse and long-term dynamical evolution of these systems is analysed using three-dimensional numerical simulations. The calculations focus on the final collapse stages of the inner baryonic component and therefore ignore the presence of dark matter. Two types of initial conditions are considered: uniformly rotating spherical clouds, and irrotational ellipsoidal clouds. In both cases the clouds are initially cold, homogeneous, and not far from rotational support ($T/|W| \approx 0.1$). Although the details of the dynamical evolution depend sensitively on the initial conditions, the qualitative features of the final configurations do not. Most of the gas is found to fragment into small dense clumps, that eventually make up a spheroidal component resembling a galactic bulge. About 5% of the initial mass remains in the form of a smooth disk of gas supported by rotation in the gravitational potential well of the outer spheroid. If a central seed black hole of mass $\gtrsim 10^6 M_\odot$ forms, it can grow by steady accretion from the disk and reach a typical quasar black hole mass $\sim 10^8 M_\odot$ in less than $5 \times 10^8$ yr. In the absence of a sufficiently massive seed, dynamical instabilities in a strongly self-gravitating inner region of the disk will inhibit steady accretion of gas and may prevent the immediate formation of a quasar.

---

1 Present address: Astronomy Department, Harvard University, Cambridge MA 02138.
2 Hubble Fellow.

# 1. INTRODUCTION

The earliest objects observed to exist in the universe are quasars (Schneider, Schmidt, & Gunn 1991). Their bright emission is often associated with the accretion of gas onto massive ($10^{6-10} M_\odot$) black holes. Current cosmological simulations (Katz et al. 1993) predict the formation of potential quasar sites at redshifts as high as $z = 8$. The corresponding density peaks involve massive ($\sim 10^{9-10} M_\odot$), self-gravitating clouds of cold gas that have collapsed to a scale $\lesssim 1\,\mathrm{kpc}$, below the spatial resolution of the simulations. However, it is far from clear whether this cold gas can collapse further to form a massive black hole. In particular, the angular momentum of the gas will prevent radial infall beyond the centrifugal barrier, at a radius $\sim 0.1$–$1\,\mathrm{kpc}$. This is still larger than the Schwarzschild radius of the system by about 6–8 orders of magnitude. The dynamics of the collapse is further complicated by small-scale fragmentation, and possibly also star formation and supernova explosions. In this paper, we study the three-dimensional hydrodynamics of the collapse and fragmentation of these rotating protogalactic gas clouds. We focus on hydrodynamics exclusively, concentrating on the possibility that massive black holes can form directly during the initial collapse of protogalaxies at high redshifts[3].

The existence of massive black holes is supported by various observational constraints on the compactness and efficiency of the energy source in active galactic nuclei. Recent observations of unresolved gravitationally lensed images (indicating a source size $\lesssim 1\,\mathrm{pc}$ from HST imaging of the quasar 2237+0305; see Rix, Schneider & Bahcall 1992), milliarcsecond jets (showing an unresolved core $\lesssim 10^{17}\,\mathrm{cm}$ in some objects; see Baath et al. 1992), gravitational microlensing (indicating a continuum source of size $\lesssim 2 \times 10^{15}$ cm in 2237+0305; see Rauch & Blandford 1991), and rapid X-ray variability (Remillard et al. 1992), all favor the black hole paradigm over its starburst alternative (Terlevich et al. 1992 and references therein). The basic question is then: How did massive black holes form at high redshifts? The answer is obviously linked to the formation of structure in the early universe (Turner 1991). The coincidence between the peak in the quasar population at $z \approx 2$ and the expected epoch of galaxy formation in the cold dark matter (CDM) cosmology may be interpreted as evidence that a quasar activity reflects the initial stage in the formation of galactic bulges (Efstathiou & Rees 1988; Rees 1992; Haehnelt & Rees 1993). Fueling of quasar black holes may also have been triggered by interactions between neighboring galactic systems (Hernquist 1989; Carlberg 1990). However, there is presently no direct observational evidence indicating whether quasar black holes formed before or after galaxies were assembled (Loeb 1993a). Relatively young galaxies in the local universe do not show evidence for the level of activity found in high-redshift quasars.

---

3 Alternatively, black holes could form as a result of the dynamical evolution of dense star clusters (Rees 1984; Quinlan & Shapiro 1990). Even in this case, the final stages of the evolution will likely involve gas dynamics.

The central theoretical issues that need to be addressed with respect to the black hole formation process are as follows:

1. **Angular momentum barrier.** Collapsing gas clouds acquire angular momentum through tidal interactions with neighboring perturbations (Hoyle 1949; Peebles 1969; Efstathiou & Jones 1979; White 1984; Hoffman 1986; Barnes & Efstathiou 1987; Ryden 1988; Quinn & Binney 1992). As the dark matter virializes, the gas cools and sinks to the center of the potential well until it becomes rotationally supported on scales of order $\sim 0.1 - 1$kpc. A black hole may form only if the specific angular momentum of the gas is reduced by 3–4 orders of magnitude. The efficient loss of angular momentum can in principle be mediated by collisional friction of free electrons or dust particles with the cosmic background radiation (Loeb 1993b; Umemura, Loeb & Turner 1993), but only for perturbations that collapse at very high redshifts ($z \gtrsim 100$). Alternatively, quasars may be associated with rare systems that aquire low values of angular momentum from tidal torques during their cosmological collapse (Eisenstein & Loeb 1993).

2. **Fragmentation and star formation.** As the gas cools and settles into a rotationally supported configuration the Jeans mass drops down to a stellar mass scale. Instead of being accreted by a central point mass the gas may then fragment into dense clumps or form stars. There is direct observational evidence that considerable fragmentation precedes quasar activity. Analysis of the broad emission lines in quasar spectra indicates a metallicity higher than solar (Hamann & Ferland 1992). An early phase of star formation and metal enrichment must therefore have occured in the quasar host systems.

3. **Formation of a single black-hole rather than a black hole cluster.** If massive black holes resulted from the growth of many (e.g. stellar mass) seeds by accretion then black hole clusters were more likely to appear rather than single black holes. A black-hole cluster forming on the scale of a galactic bulge would extend much beyond the radius of the broad line region ($\sim 10^4 GM/c^2 = 0.1\text{pc}(M/10^8 M_\odot)$; cf. Peterson 1993) and would be in conflict with constraints on the compactness of the continuum source in quasars. The growth of massive black holes in bulges must therefore start from a limited number of seeds.

For a primordial power spectrum of density fluctuations $P(k) \propto k^n$ with $n > -3$, structure forms first on small scales and black holes could have formed directly near the Jeans mass ($\sim 10^6 M_\odot$). Abundant formation of black holes in many low mass systems (Gnedin & Ostriker 1992) that subsequently combine to make larger systems would result in black hole clusters. In contrast, the CDM power spectrum has a slope $n \approx 3$ at the low-mass end. The formation of black-hole clusters may then be avoided because all mass scales below $10^{10} M_\odot$ collapse at about the same time. In addition, low-mass systems have shallower potential wells out of which the gas can easily be expelled by supernovae driven winds (Haehnelt & Rees 1993).

4. **What determines the minimum black-hole mass?** The masses of central black holes in active galactic nuclei can be deduced observationally by a variety of methods: reverberation mapping of the broad-line region (Peterson 1993), X-ray variability studies (Wandel & Mushotzky 1986), Eddington limit considerations (Haehnelt & Rees 1993), continuum fit of accretion disk models (Laor 1990), and photoionization models of the emission line intensities (Netzer 1990; Wandel & Yahil 1985; Padovani, Burg & Edelson 1990). In all cases, the deduced black-hole masses are found to be $\gtrsim 10^6 M_\odot$. Since the lowest mass objects are not necessarily observed to be close to their detection threshold, this lower bound on their mass needs to be explained.

5. **Fueling the central blak hole.** Fueling of a quasar black hole requires high overdensities ($\delta\rho/\rho \gtrsim 10^6$) within a relatively massive host object ($\gtrsim 10^{10} M_\odot$ in gas) at high redshifts (Turner 1991). If a quasar radiates near the Eddington limit the radius of the fueling system should be $\sim 0.6 \,\text{kpc} (L/10^{46} \,\text{erg} \cdot \text{s}^{-1})^{1/3} (\epsilon/0.1)^{2/3}$ and its gas density must be $\sim 1.1 \times 10^8 \, (\epsilon/0.1)^{-2} M_\odot \cdot \text{kpc}^{-3}$, where $\epsilon$ is the conversion efficiency of the accreted mass into radiation and $L$ is the quasar luminosity. Both of these characteristics correspond roughly to an evolved galactic bulge at high redshifts. It is therefore natural to link the activity in quasars to the collapse and evolution of galactic bulges.

This work will address all of the above issues in the context of a simplified hydrodynamic model for the collapse of protogalactic gas clouds in the early universe. The discussion is complementary to previous studies that concentrated on the dynamics on larger scales (Efstathiou & Rees 1988; Katz 1992; Katz et al. 1993). At galactic bulge densities the cooling time of the optically thin gas is much shorter than the hydrodynamical time, and the gas temperature remains low $T_g \lesssim 10^4 \,\text{K}$. Under these conditions, pressure forces are negligible (except in shocks where they provide dissipation), and only rotation can provide support against gravity. Here we explore the hydrodynamics of protogalactic gas clouds in this regime using hybrid $N$-body and smoothed particle hydrodynamics (SPH) simulations. The numerical method and the initial conditions for the simulations are described in §2. The results are presented in §3. In §4 we discuss the possible formation of a central seed black hole and its growth by accretion from a gaseous disk formed during the collapse. Finally, §5 summarizes our main conclusions.

## 2. NUMERICAL APPROACH AND BASIC ASSUMPTIONS

### 2.1. Smoothed Particle Hydrodynamics

The smoothed particle hydrodynamics (SPH) method has been used for the calculations presented here. SPH is a Lagrangian method that was introduced specifically to deal with astrophysical problems involving self-gravitating fluids moving freely in three dimensions. The key idea is to calculate the pressure-gradient forces by kernel estimation, directly from the particle positions, rather than by finite differencing on a grid, as is done in more traditional methods. This idea was introduced originally by Lucy (1977) and Gingold & Monaghan (1977), who applied it to the calculation of dynamical instabilities in rapidly rotating stars. Since then, a wide variety of astrophysical fluid dynamics problems have been tackled using SPH (see Monaghan 1992 for a recent review). In the past few years, these have included large-scale structure formation in cosmology (Katz, Hernquist, & Weinberg 1992) galaxy formation (Katz 1992), star formation (Monaghan & Lattanzio 1991), supernova explosions (Herant & Benz 1992), and stellar interactions (Rasio & Shapiro 1991, 1992).

We use a modified version of the SPH code developed originally by Rasio (1991) for the study of hydrodynamic stellar interactions (Rasio & Shapiro 1991, 1992). The implementation of the SPH scheme is similar to that adopted by Hernquist & Katz (1989), but the gravitational field is calculated using a fast grid-based FFT solver. The neighbor searching in the code is also performed using a grid-based algorithm. Specifically, a multigrid hierarchical version of the linked-list method usually adopted in $P^3M$ particle codes (Hockney & Eastwood 1988) has been developed. The improved algorithm is extremely efficient, even for very nonuniform distributions of particles, provided that one is careful to fine-tune the ratio $L_g/\langle h_i \rangle_g$ of grid separation $L_g$ to the average SPH smoothing length $\langle h_i \rangle_g$ for that grid. Other details about the implementation, as well as the results of a number of test-bed calculations were presented in the papers cited above.

The most important modification made to the code for this study is the addition of a particle-mesh (PM) integrator that allows us to treat a collisionless component in the system, in addition to the fluid component. The collisionless particles are treated basically like SPH particles, except that only the gravitational force is retained in their equation of motion. A simple cloud-in-cell scheme (see, e.g., Hockney & Eastwood 1988) is used for both assigning density to, and interpolating forces from grid points. Thus the code can now be used for hybrid PM/SPH simulations. In the calculations of dissipational collapse described here, the PM method is used to model the system of compact high-density clumps of gas that form due to the gravitational instability. Specifically, whenever the density $\rho_i$ at the position of an SPH particle becomes larger than some critical value, the SPH particle is transformed instantaneously into a collisionless particle. The gas represented by that SPH particle is assumed to continue its collapse to smaller and smaller scales, so that it effectively decouples from the rest of the fluid, except for the gravitational interaction. This procedure, although admittedly fairly crude, allows us to model the dynamical evolution of a collapsing system well past the point where the rapidly increasing dynamic range in fluid densities would make a purely hydrodynamic calculation grind to a halt (because the timestep must satisfy $\Delta t \lesssim (G\rho_{\max})^{-1/2}$). The feedback effects of star formation on the large-scale hydrodynamics is ignored. Indeed, in the galaxy formation simulations of Katz (1992), which incorporated crudely the effects of star formation and energy production in supernovae, this feedback was found to have little effect on the numerical results.

Another important modification is that the SPH energy (or entropy) equation appropriate for adiabatic conditions has been replaced by an isothermal equation of state. This is equivalent to setting the adiabatic exponent $\Gamma = 1$ and letting the "entropy variables" $A_i \equiv p_i/\rho_i^\Gamma = c_s^2 = k_B T_g/\mu$ be constant for all particles (where $p_i$ is the pressure and $\mu$ the mean molecular weight). The assumption of a constant temperature is amply justified in our model, since the gas is optically thin and cooling is so efficient that the temperature remains always $\lesssim 10^4$ K. The exact value of the temperature is unimportant, since pressure-gradient forces remain always negligible during the collapse. Shock dissipation is provided by the standard artificial viscosity, with the quadratic (von Neumann-Richtmyer) term dominating in this isothermal regime.

All calculations shown here were done using $N \approx 10^4$ particles (total, both SPH and collisionless), and each SPH particle interacting with a nearly constant number of neighbors $N_N \approx 64$. The gravitational potential is calculated by FFT on a $256^3$ grid. A typical run takes about 10 CPU hours on an IBM ES-9000 supercomputer.

## 2.2. Initial Conditions

For simplicity we consider an initially homogeneous cloud, represented by a random spatial distribution of SPH particles all having the same individual mass. We do not introduce a spectrum of small-scale density fluctuations, and avoid relaxing the initial distribution of SPH particles, since this would introduce strong correlations between particle positions on scales comparable to the particle smoothing lengths $h_i$. We adopt units such that $G = M = R = 1$ where $M \equiv M_9 \times 10^9 M_\odot$ is the total mass and $R \equiv R_3$ kpc is the initial size of the system. In these units the hydrodynamical time $t_{\rm dyn} \equiv (R^3/GM)^{1/2} \approx 10^7 \, {\rm yr} \, R_3^{3/2} M_9^{-1/2}$. We assume an isothermal equation of state $P_i = c_s^2 \rho_i$ with $c_s^2 = 0.01$ in our units, corresponding to a temperature $T_g \approx 5 \times 10^3 \, {\rm K} \, M_9 R_3^{-1} (\mu/m_p)$, where $m_p$ is the proton mass. This is justified since the cooling time of the gas $t_{\rm cool} \lesssim 10^5$ yr remains much smaller than the dynamical time $t_{\rm dyn}$ throughout the evolution.

Angular momentum in the system is added in one of two ways. The first type of initial condition we consider is a *rigidly rotating* spherical cloud. The angular velocity is given by $\Omega^2 = 0.3$ in our units, corresponding to a ratio of rotational kinetic energy to gravitational binding energy $T/|W| = \frac{1}{3}\Omega^2/(GM/R^3) = 0.1$ initially. The total angular momentum is then given by $J/(GM^3R)^{1/2} \approx 0.2$.

The second type of initial condition we consider is an *irrotational* (zero-vorticity) *ellipsoidal* cloud. To produce angular momentum without vorticity, we use the velocity field of a classical irrotational Riemann ellipsoid of type S (Chandrasekhar 1969, Chap. 7). Riemann ellipsoids are *equilibrium* fluid configurations supported by a combination of solid body rotation and internal fluid motions of *uniform* vorticity. In the type-S ellipsoids, the vorticity vector is everywhere parallel to the rotation axis. The irrotational ellipsoids are a particular case where the solid body rotation and internal motions combine to produce zero vorticity as seen in the inertial frame. Our initial condition is of course not in equilibrium, but the velocity field is the same as in the equilibrium solution. In addition, we adjust the axis ratios of the ellipsoidal cloud so that the (uniform) density and total angular momentum are exactly the same as in the uniformly rotating sphere considered previously. The motivation here is that precisely this type of irrotational velocity field tends to be produced by external tidal torques acting on a mass of fluid (see, e.g., Kochanek 1992). Vorticity cannot be produced by tidal torques in the absence of dissipation. Thus, an irrotational configuration is most likely if the angular momentum of the baryonic component in the protogalaxy was induced by external tidal torques before the collapse became dissipational. If, on the other hand, orbit crossing and dissipative interactions (e.g., with a dark matter halo; see Katz 1992) have already occurred, then the fluid velocity field may have acquired vorticity and more organized fluid motions resembling solid body rotation are possible. The two types of initial conditions considered in this paper are meant to represent these two different regimes.

## 3. NUMERICAL RESULTS

### 3.1. *Collapse of a Uniformly Rotating Sphere*

The initial phase of the dynamical evolution (Fig. 1a, b) is characterized by the overall collapse of the system to a thin disk and the formation of dense filaments and clumps of cold gas. In the absense of an appropriate treatment, this process of continued fragmentation to smaller and smaller scales would make the numerical calculation grind to a halt well before a final equilibrium state is reached. This is because the hydrodynamic time inside clumps rapidly becomes orders of magnitude shorter than that of the system as a whole. As explained in §2, we solve this problem by converting gas particles into point-like objects whenever the local density at the particle's position is above a critical value $\rho_{\max}$. These point-like objects, representing high-density gas clumps at the limit of the numerical resolution, are treated subsequently as a collisionless system interacting with the remaining gas though gravitational forces only. In the simulation of Fig. 1, we used $\rho_{\max}/\bar{\rho} = 10^6$, where $\bar{\rho}$ is the mean mass density in the system. This choice of $\rho_{\max}$ is dictated primarily by practical considerations, since the total computation time of each simulation increases like $\rho_{\max}^{1/2}$. Calculations performed with different values of $\rho_{\max}/\bar{\rho}$ in the range $10^4$–$10^6$ show small differences in the details of the evolution, but remain all in good qualitative agreement.

In just a few dynamical times, the dense gas clumps and filaments collide and merge, leading to fragmentation on smaller and smaller scales until a significant fraction, typically ≈95%, of the total mass is converted into the collisionless component (Fig. 2). At intermediate times (Fig. 1c, d), a tightly bound, compact group containing a small number ($\lesssim 10$) of individual subsystems is observed. Further interactions between these subsystems lead to mergers, so that only two subsystems remain after a few crossing times (Fig. 1e), and these finally merge to produce the single structure obtained at the end of the calculation (Fig. 1f). This final, quasi-equilibrium configuration consists of a thin disk of cold gas embedded in a virialized, collisionless halo of a spheroidal shape[4]. The gas is supported by rotation in the external halo potential. No measurable deviation from axisymmetry is left in this final configuration, although some evidence for the formation of a bar in the innermost region of the gaseous disk may be identified in Figure 1f (and is expected theoretically, cf. §4.1 below). We did not attempt to follow numerically the subsequent dynamical evolution of the gas in this inner region, since it is at the limit of our spatial resolution.

### 3.2. *Collapse of an Irrotational Ellipsoid*

The initial phase of the collapse of an irrotational ellipsoidal cloud is strikingly different (Fig. 3a, b). Here the system as a whole collapses to a *spindle-like configuration* at first. However, fragmentation on small scales and the formation of high-density gas clumps proceed as before. After several dynamical times, a small number of orbiting subsystems are again seen, although the overall mass distribution appears more centrally concentrated (Fig. 3c, d). The subsystems quickly merge into a single configuration (Fig. 3e,f): a virialized halo of dense gas clumps forms, while a small fraction $\approx 4\%$ of the mass remains in a thin gaseous disk. Thus, *quite independent of the details of the initial conditions*, it appears that the late stages of the dynamical evolution are always dominated by violent relaxation of the high-density clumps, and the collisions and mergers between a small number of subsystems. These complex dynamical processes tend to erase any memory of the initial conditions, and ultimately determine the principal characteristics of the final configuration.

---

4 It is interesting to note that a similar structure is seen in the bulge of our own Galaxy, where radio observations show the presence of a cold disk containing $\sim 10^8 M_\odot$ of molecular gas (i.e., a few percent of the total bulge mass) extending out to a distance $\sim 300\,\text{pc}$ from the Galactic center (Blitz et al. 1993).

## 4. DISCUSSION

### 4.1. *Disk Evolution*

It is tempting to envision that viscous accretion of the $\sim 10^8 \, M_\odot$ of gas in the central disk onto a central seed object could lead to the formation of a supermassive black hole. This would occur on a viscous timescale,

$$t_{\rm vis} \sim 5 \times 10^8 \text{ yr } M_9^{1/2} R_3^{1/2} \alpha^{-1} (c_s/10 \text{ km s}^{-1})^{-2}, \tag{1}$$

where $c_s$ is the sound speed in the disk, and $\alpha \lesssim 1$ is a dimensionless measure of effective viscosity (Shakura & Sunyaev 1973). As the disk shrinks in size it may reach a point where its further contraction is limited by cooling rather than by angular momentum transport. In the process of losing angular momentum the disk radiates its gravitational binding energy with a luminosity, $L \sim GM^2/Rt_{\rm vis}$. This luminosity would approach the Eddington limit at a radius $\sim 0.1 \text{pc} \times M_9^{1/3} \alpha^{2/3} (c_s/10 \text{km} \cdot \text{s}^{-1})^{4/3}$. The subsequent evolution would still be limited by a relatively short cooling timescale,

$$t_{\rm cool} \sim \epsilon t_E < 4.4 \times 10^8 \text{yr}, \tag{2}$$

where $\epsilon$ is the overall efficiency for convering rest mass into radiation, and $t_E \equiv Mc^2/L_E = 4.4 \times 10^8$ yr is the Eddington time.

However, even in the presence of a stabilizing external potential, an initially cold gaseous disk is prone to numerous instabilities (Binney & Tremaine 1987). Local stability to axisymmetric perturbations requires that the Toomre ratio $Q \equiv c_s \kappa/(\pi G \Sigma) > 1$, where $\Sigma \sim \zeta M/(\pi R^2)$ is the disk surface density ($\zeta \sim 0.1$ is the ratio of disk to halo mass) and $\kappa \sim 2(GM/R^3)^{1/2}$ is the epicyclic frequency. For the present system we have $Q \sim 3(\zeta/0.1) M_9^{-1/2} R_3^{1/2} (c_s/10 \text{ km s}^{-1})$, indicating marginal stability. Depending on the poorly known details of the thermal processes involved, some disks could easily become locally unstable everywhere. In such a case we expect further fragmentation of the disk material into cold gas clouds or stars (as observed in the disks of present spiral galaxies), and the growth of a central supermassive black hole by viscous accretion would not take place.

Even if local stability is achieved, a self-gravitating disk can still be unstable to global, non–axisymmetric modes. In the absence of a central compact mass, the self-gravity of the disk (whose mass interior to $r$ scales like $r^2$ close to the center) will always become dominant over the external halo gravity (mass $\propto r^3$). In that case the central region of the disk would behave like an isolated, self-gravitating disk of cold gas, which is known to be violently unstable to non–axisymmetric modes (Binney & Tremaine 1987). To obtain global stability, we require the presence of a central seed mass $M_s$ such that for all $r$,

$$G\Sigma < \frac{4\pi}{3}G\rho_h r + \frac{GM_s}{r^2}, \qquad (3)$$

where the halo density $\rho_h \sim M/(4\pi R^3/3)$. With $\Sigma \sim \zeta M/(\pi R^2)$ the stability condition (3) translates to the constraint

$$M_s > 10^6 \ M_\odot \times M_9 \left(\frac{\zeta}{0.1}\right)^3. \qquad (4)$$

For $\zeta \approx 0.1$, the above inequality is consistent with the fact that the empirically deduced black hole masses are always $\gtrsim 10^6 M_\odot$ (cf. point 4 in the introduction).

The presence of a sufficiently massive central seed object is essential for the growth of a massive black hole near the center of the gaseous disk. One possibility is that the central seed forms during the initial collapse out of gas with low angular momentum (Eisenstein & Loeb 1993). As illustrated in Figure 4, there are two routes that could lead to the formation of a seed black hole out of this gas. If the cooling time of the gas is shorter than its viscous time (cf. Eqs. [1] & [2]), the gas will settle into a self-gravitating supermassive disk, supported mainly by rotation (cf. §4.2). Angular momentum transport by nonaxisymmetric instabilities or by viscosity could eventually lead to the formation of a seed black hole for a sufficiently compact initial disk. On the other hand, if the viscous time is shorter than the cooling time the gaseous cloud could collapse quasi-spherically. The collapse would either lead directly to a black-hole (e.g., via radiative viscosity; cf. Loeb & Laor 1992) or could be slowed down by radiation pressure if the entropy is high enough. In this latter case, a supermassive star would form as an intermediate state (Wagoner 1969; Shapiro & Teukolsky 1983). The dominance of radiation pressure in both supermassive stars and compact supermassive disks would prevent fragmentation and star formation in these systems.

### 4.2. Supermassive Stars and Disks

It is possible that during the initial collapse a small amount of gas with very low specific angular momentum will condense near the center of the potential well and collapse quasi-spherically. As its density rises, the gas must eventually become optically thick. When the photon diffusion time becomes longer than the gas infall time, radiative cooling becomes inefficient and the gas heats up adiabatically, with the pressure rising as $p \propto \rho^\Gamma$. If radiation pressure becomes dominant during the collapse, the adiabatic index of the gas will approach the value $\Gamma = 4/3$, and pressure support will never be able to halt the collapse since $\partial p/\partial r \propto 1/r^5 \propto GM\rho/r^2$. However, if the entropy production in shocks is sufficiently large, significant pressure support may build up, leading to the formation of a *supermassive star* (Hoyle & Fowler 1963)[5].

In most cases, however, rotation is likely to affect the collapse earlier than the pressure support. Without loss of angular momentum the rotational velocity of a fluid element scales as $v_\phi \propto r^{-1}$, while its radial infall velocity rises only as $r^{-1/2}$. Therefore a supermassive star of radius $R_{\rm sm}$ can form out of a cloud with an initial radius $R_0$ only if the initial ratio of rotational to radial velocities is $\ll (R_{\rm sm}/R_0)^{1/2}$. While this constraint may be satified in rare cases (Eisenstein & Loeb 1993), a majority of systems must lose angular momentum very efficiently for the final stage of the collapse to be quasi-spherical. In the absence of an efficient mechanism for removing angular momentum, a *supermassive disk* (Wagoner 1969) must form.

It is usually assumed that supermassive stars are convective, and can therefore be modeled as polytropes with an adiabatic index $\Gamma \approx 4/3$ (Wagoner 1969; Shapiro & Teukolsky 1983). Since we could not find a justification for this assumption in the literature, we provide a simple but rigorous proof in appendix. Following equations $(A2)$ and $(A3)$, the star radiates at the Eddington luminosity, $L_E \equiv 4\pi GMm_p/\sigma_T = 1.3 \times 10^{46}$ erg s$^{-1}$ $(M/10^8 M_\odot)$. Without rotation the star would contract on a timescale much shorter than the Eddington cooling time, $t_E \equiv Mc^2/L_E = 4.4 \times 10^8$ yr, until it gets to a radius where the general relativistic instability sets in (Shapiro & Teukolsky 1983). A rotating star must shed mass and angular momentum in order to reach this state. Because of the mass loss, and since rotation has a stabilizing effect, the radius of the star at the onset of gravitational collapse will be smaller than the value $1600(GM/c^2) \times (M/10^6 M_\odot)^{1/2}$ obtained in the nonrotating case (Shapiro & Teukolsky 1983). For a *uniformly rotating* supermassive star, the mass shedding limit is reached at a very small value of the ratio of rotational kinetic energy to gravitational binding energy, $(T/|W|)_{\rm max} \approx 0.007$ (Zel'dovich & Novikov 1971). In reality, however, the evolution of a rotating supermassive star will depend sensitively on the internal distribution of specific angular momentum. Even a small amount of differential rotation could stabilize the star against mass shedding, allowing much highers values of $T/|W|$ to be reached (see Bodenheimer & Ostriker 1973).

---

[5] The question of entropy production in spherical collapse will be addressed in detail elsewhere (Thoul, Rasio & Loeb 1993).

Under the assumption that convection leads not only to uniform specific entropy, but also to uniform rotation[6], Bisnovati-Kogan, Zel'dovich & Novikov (1967) have calculated the evolution of a rotating supermassive star, modelled as an $n = 3$ polytrope. Once the contracting star reaches the mass-shedding limit, an equatorial outflow must appear. The subsequent evolution was modeled as a sequence of maximally rotating ($T/|W| = (T/|W|)_{\max}$ =constant) polytropic configurations with decreasing mass and angular momentum. It was found that rotational support increases the lifetime of the supermassive star by up to a factor of 10 compared to the nonrotating case. Depending on the initial value of the angular momentum, between about 1% and 30% of the total mass is lost to an equatorial disk before the central star becomes unstable and collapses to a black hole. These results are for an object of mass $10^9 M_\odot$, the only value considered by Bisnovatyi-Kogan et al. (1967). For a smaller star, it is possible that most of the mass will be lost to an outer disk. Therefore, even a very slowly rotating supermassive star could transform itself into a supermassive disk before becoming unstable to quasi-spherical collapse.

Clearly, supermassive disks seem much more likely to form than supermassive stars. Unfortunately, both the dynamical properties and the secular (viscous and thermal) evolution of isolated self-gravitating disks are not well understood theoretically. They depend sensitively on the unknown mass and angular momentum distributions, on the physical origin of the effective viscosity (which affects the heating rate and mass transfer rate as a function of radius and time), and on the transport of mass and angular momentum by dynamical instabilities. The secular evolution of isolated massive disks was discussed recently by Eisenstein & Loeb (1993), assuming a simple $\alpha$-viscosity. The viscous time was found to be shorter than $10^{6-7}$yr (comparable to the star formation time) for the initial collapse of rare, low-angular-momentum objects in the universe. Most of the seed black holes were found to form just above the cosmological Jeans mass $\sim 10^6 M_\odot$ (with roughly one seed object per bright galaxy), which, by coincidence, satisfies the constraint of equation (4).

---

6 This assumption is apparently not satisfied in the convective envelope of the sun, where helioseimological data shows evidence for differential rotation (Brown et al. 1989). The solar rotation profile can be linked to the kinetics of convective turbulence (Kumar, Loeb & Narayan 1993).

## 5. SUMMARY

We have performed numerical simulations of the hydrodynamic collapse of rotating clouds of cold gas in the centers of protogalaxies. We considered two simple limiting cases for incorporating angular momentum in the initial configuration. The first is solid body rotation, representing the well-organized fluid motions that could result from earlier dissipational evolution. The second is an irrotational velocity field, of the type produced by dissipationless tidal interactions with other distant masses. In both cases we find that most of the gas fragments into dense clumps. We model these clumps as a collisionless system that interacts with the remaining gas only through gravity. During the evolution, a few subsystems form and then rapidly collide and merge together into a single virialized structure resembling the bulge of a spiral galaxy. The remaining gas forms a smooth thin disk that contains a fraction $\zeta \lesssim 0.1$ of the total mass and is supported by rotation in the external potential of the collisionless component. A central seed black hole of mass $\gtrsim 10^6 M_\odot \times M_9(\zeta/0.1)^3$ is necessary in order to stabilize the central region of the gaseous disk, allowing the smooth growth of the central seed by accretion. This minimum seed mass is consistent with the lowest values of the empirically determined black-hole masses in active galactic nuclei. A seed black hole could form directly through the initial collapse of gas with very low specific angular momentum. In some cases, a quasi-spherical supermassive star may form as an intermediate state. More commonly, rotational support will precede pressure support and lead to the formation of a supermassive disk as an intermediate state.

This work has been supported in part by a W. M. Keck Foundation Fellowship to A. L., and by a Hubble Fellowship to F. A. R., funded by NASA through Grant HF-1037.01-92A from the Space Telescope Science Institute, which is operated by the Association of Universities for Research in Astronomy, Inc., under contract NAS5-26555. F. A. R. also acknowledges gratefully the hospitality of the ITP at UC Santa Barbara, where part of this paper was written. Computations were performed on the Cornell National Supercomputer Facility, a resource of the Center for Theory and Simulation in Science and Engineering at Cornell University, which receives major funding from the NSF and IBM Corporation, with additional support from New York State and members of its Corporate Research Institute.

# Appendix

In the following we show that the radiative cooling of a supermassive star inevitably leads to the development of convection. Consider a supermassive star which is purely radiative initially, i.e. has a stable entropy profile. Without a nuclear energy source (nuclear burning can be ignored for masses $\gtrsim 10^6 M_\odot$; cf. Wagoner 1969), the entropy of each electron-proton fluid element must be changing according to the local radiative heat flux $\vec{F}$ at a rate given by

$$\frac{2T_g}{m_p}\frac{\partial s}{\partial t} = -\frac{1}{\rho}\vec{\nabla}\cdot\vec{F}, \tag{A1}$$

where $\rho$, $T_g$ and $s$ are the mass density, temperature and specific entropy, and $m_p$ is the proton mass. If the opacity is dominated by Thomson scattering, the local radiative heat flux $\vec{F}$ is related to the radiation pressure gradient by

$$\vec{F} = -\frac{m_p}{\sigma_T}\frac{1}{\rho}\vec{\nabla}p_\gamma, \tag{A2}$$

where $\sigma_T$ is the Thomson cross-section. Ignoring gas pressure and rotation, the hydrostatic equilibrium equation is simply

$$\frac{1}{\rho}\vec{\nabla}p_\gamma = \vec{g}, \tag{A3}$$

where the gravitational field $\vec{g}$ obeys Poisson's equation,

$$\vec{\nabla}\cdot\vec{g} = -4\pi G\rho. \tag{A4}$$

Combining equations (A2)–(A4), we see that the RHS of equation (A1) is constant,

$$\frac{1}{\rho}\vec{\nabla}\cdot\vec{F} = \frac{4\pi G m_p}{\sigma_T} = \text{constant}. \tag{A5}$$

By taking the gradient of equation (A1) we then obtain

$$\frac{\partial}{\partial t}\vec{\nabla}s = \frac{2\pi G m_p^2}{\sigma_T}\frac{\vec{\nabla}T_g}{T_g^2} < 0. \tag{A6}$$

The temperature gradient in the radial direction is negative since heat flows out of the star. Therefore, a negative radial entropy gradient must develop eventually, leading to convection. For a rotating supermassive star, the above result holds as long as the rotation period is much longer than $(G\rho)^{-1/2}$.

# FIGURE CAPTIONS:

Fig. 1: Dynamical evolution of a spherical, uniformly rotating cloud. Gas particles are shown on the right, collisionless (high-density) particles on the left. The units are defined in §2.

Fig. 2: Time evolution of the fractional mass in the smooth gaseous component during the dynamical collapse of a uniformly rotating spherical cloud (solid line) and an irrotational ellipsoidal cloud (dashed line).

Fig. 3: Dynamical evolution of an ellipsoidal, irrotational cloud. Conventions are as in Fig. 1.

Fig. 4: Possible evolutionary paths leading to the formation of a central seed black hole. As shown in Figs. 1 and 3, the initial collapse of a protogalactic gas cloud results in the formation of a thin disk supported by rotation in the external potential of a collisionless, bulge-like component. The inner region of the disk is initially unstable and may become optically thick. If the viscous time for angular momentum transport is shorter than the cooling time, the gas will become radiation-pressure supported and could form a supermassive star. The star contracts quasi-spherically and spins up as it radiates at the Eddington rate $L_E$. It becomes unstable to mass shedding when the ratio $T/|W| \lesssim 0.01$. If most of its mass is not lost to an outer disk, the star will ultimately become gravitationally unstable and collapse to a black hole. If, initially, the viscous time is longer than the cooling time, a supermassive disk would form. Such a self-gravitating disk is unstable to nonaxisymmetric instabilities when $T/|W| \approx 0.3$. Through various sources of effective viscosity, the central part of the disk will lose angular momentum and may evolve directly into a black hole on the viscous timescale.